\documentclass[aps,prab,reprint,superscriptaddress,longbibliography,floatfix]{revtex4-2}

\usepackage{graphicx}
\usepackage{amsmath,amssymb}
\usepackage{xcolor}
\usepackage[colorlinks=true,linkcolor=blue,citecolor=blue,urlcolor=blue]{hyperref}

\newcommand{\tabench}{Ta\_bench}
\graphicspath{{figs/}}

\begin{document}

\title{Experimental verification of the PRIORA Monte Carlo code for proton radiography}

\author{D.~Varentsov}
\email{d.varentsov@gsi.de}
\affiliation{GSI Helmholtzzentrum f\"ur Schwerionenforschung, Darmstadt, Germany}
\author{M.~Schanz}
\affiliation{GSI Helmholtzzentrum f\"ur Schwerionenforschung, Darmstadt, Germany}
\author{J.~Schmidt}
\affiliation{Los Alamos National Laboratory, Los Alamos, NM, USA}

\date{27 October 2026}

\begin{abstract}
PRIORA (Proton Radiography Ion-Optical Ray-tracing Analysis) is a fast
Monte Carlo model for high-energy proton radiography. It combines fitted
proton--matter interaction tables with energy-dependent transfer maps of
arbitrary order and simulates approximately $10^6$ protons/s through a
complete ion-optical system on an 8-core laptop. We verify it with GSI data obtained
with the \tabench{} target: fifteen precision slits in a 2\,mm Ta plate
between symmetric PMMA staircases. A single experimental image provides
the image scale, spatial resolution as a function of chromatic load,
the direction of longitudinal defocus in both transverse planes,
effective angular acceptance and
multiple-scattering contrast. With identical
processing of experiment and simulation and no physics adjustment,
PRIORA reproduces the data obtained with 2 and 4\,GeV protons. At
2\,GeV, PRIORA also recovers sub-percent quadrupole trims consistent
with the control-system records and determines a target-plane spatial
resolution of $115\pm6\,\mu$m.
\end{abstract}

\maketitle

\section{Introduction}
\label{sec:intro}

Image-level proton-radiography simulations require two coupled models:
proton--matter transport in thick structured objects and high-order
chromatic optics of the magnetic microscope. General-purpose transport
codes (Geant4, FLUKA and MCNP~\cite{Geant4,FLUKA,MCNP}) describe
scattering, energy loss, straggling and nuclear interactions; modelling a
particular magnetic microscope additionally requires its field and
geometry description. Map codes such as COSY Infinity~\cite{COSY} contain the optics,
but not material image formation. The required result is a complete
radiograph with $10^9$--$10^{10}$ protons, finite collimation and the
actual microscope state.

PRIORA (Proton Radiography Ion-Optical Ray-tracing Analysis) retains
Monte Carlo image formation, but replaces microscopic interaction
tracking within each material slice by sampling an effective material
response. Its scattering, displacement, energy-loss, straggling and
removal parameters are fitted to LADA calculations. LADA (Laser and
Accelerator Differential Algebra) is a particle-transport and ion-optics
code used here to calculate material responses and generate high-order
transfer maps~\cite{LADA}. The optics is
represented by energy-dependent transfer maps, third order in this
study. Matter and optical elements alternate, so each map acts on the
actual particle energy. The finite W collimator is treated as matter with
its machined aperture, and the surviving particles are histogrammed at
the detector. The instrument state can therefore be fitted together with
the image instead of being imposed as an external assumption.

The \tabench{} target was chosen for the experimental verification of
PRIORA. A single experimental radiograph provides absolute scale,
spatial resolution as a function of chromatic load, the direction of
longitudinal defocus in both transverse planes, effective angular
acceptance and multiple-scattering contrast.
A single reconstructed instrument state must reproduce all these
measurements simultaneously.

\section{The PRIORA model}
\label{sec:priora}

\subsection{Architecture}

The PRIORA model is a ``train'' of a random beam generator, matter
targets, ion-optical transfer maps of arbitrary order and an image saver.

\paragraph{Matter.} For each material and reference energy, PRIORA uses
functions of path length $L$ and relative energy deviation $k$: the rms
scattering angle $\sigma_a(L,k)$, the mean kinetic-energy loss
$\Delta K(L,k)$ and the energy-loss straggling $\sigma_K(L,k)$, together
with nuclear removal calculated from tabulated collision lengths. These
functions are obtained from full LADA matter-transport
calculations~\cite{LADA}, including stopping, multiple scattering,
energy-loss straggling and nuclear removal.

Thick material slices are automatically divided into thinner
sub-slices before transport. Within each sub-slice of thickness $\ell$,
a scattering kick $\theta$ at longitudinal coordinate $z$ is accompanied
by the correlated transverse displacement $\theta(z-\ell/2)$ relative
to the sub-slice mid-plane. Energy loss, straggling and nuclear removal
are also applied separately in each sub-slice. This preserves both the
energy evolution through the object and the longitudinal position of
each scattering event required for the focus diagnostics below.

\begin{figure*}[tb]
  \centering
  \includegraphics[width=\textwidth]{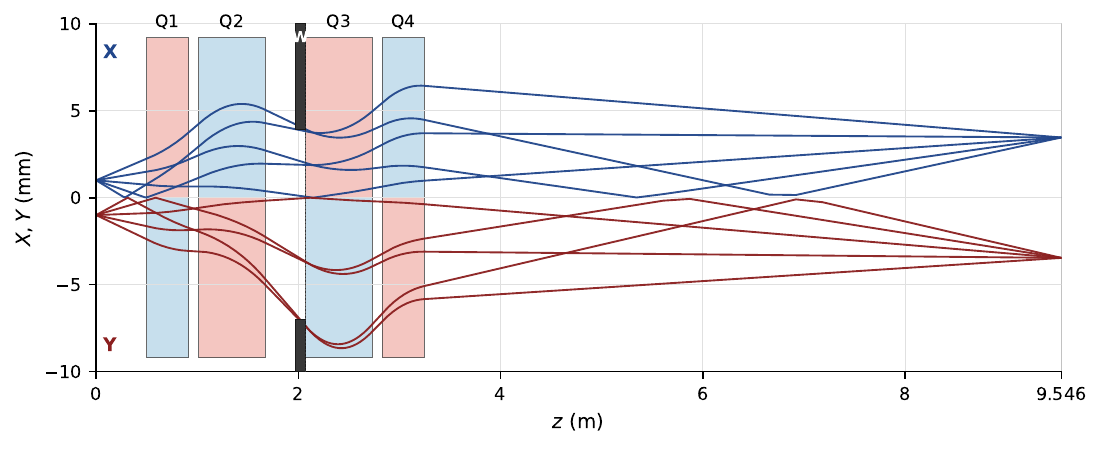}
  \caption{LADA third-order X/Y ray trace of the PRIOR-II proton microscope
  using measured magnet fields and the fixed geometry of
  Ref.~\cite{Schanz2024}. The 100-mm W collimator is located immediately
  upstream of Q3 at the Fourier plane. The rays form a point-to-point image
  at $z=9.546$~m.}
  \label{fig:scheme}
\end{figure*}

\paragraph{Optics.} The ion-optical system is represented by transfer
maps of arbitrary order, generated by LADA from measured magnet-gradient
profiles~\cite{Schanz2024} or imported from COSY~\cite{COSY}. Matter and
maps alternate in the train, so energy loss inside an extended object
changes the subsequent ion-optical transport automatically. In PRIOR-II
(Fig.~\ref{fig:scheme}), the Q1--Q4 magnetic quadrupoles image the object
with $M\approx-3.5$. The Fourier-plane collimator selects protons
according to the scattering angle in the object and thereby determines
the material contrast.

\paragraph{Collimation and detection.} Fourier-plane collimators are
finite targets, here 100\,mm W with the machined elliptical aperture, not
ideal cuts. Edge scattering and partial transmission are therefore
included. Scintillators can be appended as detector-plane matter; the
image is a configurable two-dimensional histogram.

\subsection{Performance}
\label{sec:fast}

High throughput is achieved by evaluating fitted material-response
functions on vectorized particle arrays and processing independently
seeded chunks of random particles in parallel. For the full PRIOR-II
model used here (matching line, 132\,mm three-material target, four
quadrupoles and finite W collimator~\cite{Schanz2024}), the calculation
rate on an 8-core laptop is approximately $10^6$ protons/s. A
$5\times10^8$-proton radiograph takes about 8 minutes; a target/empty
pair with $2\times10^9$ protons per image takes about one hour.
A target/empty pair comprises two simulated images with identical beam
and microscope settings, with and without the object. The simulation
and analysis scripts automate both calculations and form normalized
target/empty ratios.

\subsection{Analysis conventions}
\label{sec:conventions}

All comparisons are expressed in target-plane coordinates. Experimental
and simulated images are registered separately to the known slit
geometry, normalized to the open-slit intensity, and the simulation is
not smoothed. Each simulated image is first normalized by its total
intensity. For the PMMA-band profiles, target and empty intensities are
integrated over the same ROI before division; the resulting profiles are
then normalized to the open-slit level. Each slit profile contains a
rising and a falling edge.
Both edges are fitted simultaneously with error-function edge-spread
functions (ESFs) having a common Gaussian width $\sigma$. The difference
between the fitted edge positions gives the slit width, whereas the
common Gaussian width determines the spatial resolution. The reported
resolution is $2\sqrt{2\ln2}\,\sigma$, the FWHM of the corresponding
Gaussian line-spread function, i.e. the ESF derivative. The same fit is
applied to experiment and simulation. Any detector response is applied
identically to the target and \textquotedblleft empty\textquotedblright{}
images before division.

\section{The \tabench{} target}
\label{sec:target}

\begin{figure}[tb]
  \centering
  \includegraphics[width=0.86\linewidth]{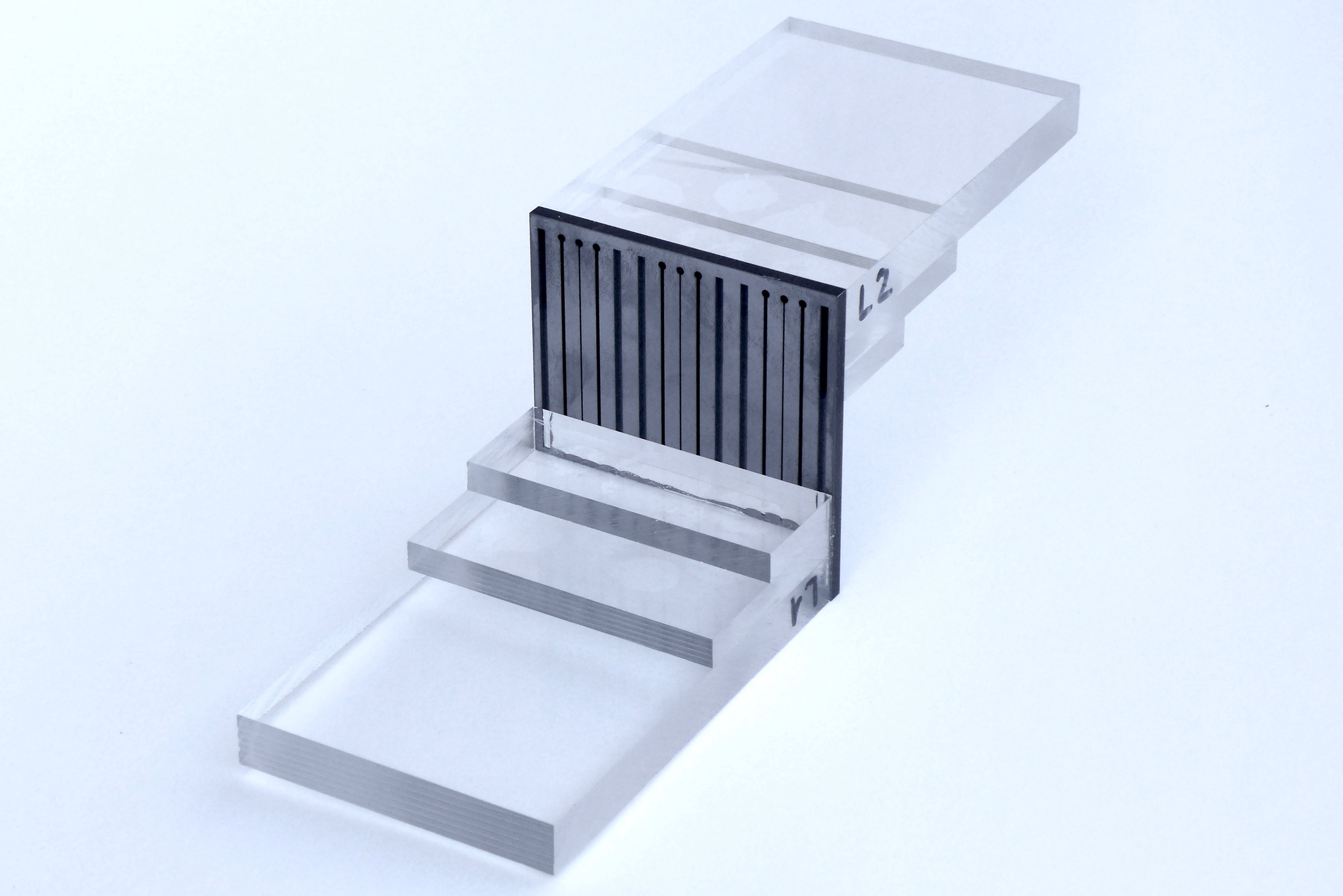}
  \caption{The \tabench{} target: the 2\,mm Ta slit plate between
  the two symmetric monolithic PMMA staircases (20, 45 and 65\,mm
  cumulative thickness). The fifteen through-slits form three groups of
  0.5/0.375/0.25/0.375/0.5\,mm width over a 20\,mm baseline.}
  \label{fig:target}
\end{figure}

The target gives six observables in one exposure. Slit positions give
scale and magnification. Slit ESF widths in the PMMA bands give
resolution versus controlled energy loss. The pre/post asymmetry
identifies whether the object-plane focus lies upstream or downstream
of the Ta plate in the slit-sensitive plane. Upstream scattering acts
with the lever
$\theta dz$, downstream scattering with $\theta(dz+z)$, where $z$ is the
distance from the plate to the PMMA scatterer. The sign of the blur
asymmetry is therefore the sign of the longitudinal focus error. The
PMMA staircase boundaries are edges in the orthogonal plane and give the
second focus sign by the same construction. The Ta-plate/open-slit
transmission ratio tests multiple scattering through the finite collimator
acceptance. The 0.25\,mm slits probe image contrast near the resolution limit.

All observables are differential. Scale does not require a camera
calibration; focus signs do not require an absolute resolution; contrast
does not require absolute flux. Together, these observables constrain
magnification, focus in both transverse planes, chromatic blur and
collimator acceptance. Because several observables depend on the same
instrument parameters, a reconstructed state must describe all PMMA
bands, slit groups and both transverse planes simultaneously.

\section{Comparison with measured absolute observables}
\label{sec:tier2}

With the design state and no adjustable parameters, the 2\,GeV
\tabench{} data (3.0\,mrad collimator, CsI/PCO Edge 5.5 camera) already
fix the absolute scale and contrast. The slit scale agrees with the map
prediction to better than 1\% and with the 4\,GeV data to 0.6\%. The
Ta-plate/open-slit transmission ratio is 0.31--0.34; the model gives 0.29.
Open-slit PMMA transmissions are $1.0/0.85/0.60$ (no PMMA/20/45\,mm);
attenuation plus angular cut gives $1.0/0.88/0.62$.

\section{Reconstruction of the instrument state}
\label{sec:tier3}

\begin{figure}[tb]
  \centering
  \includegraphics[width=\linewidth]{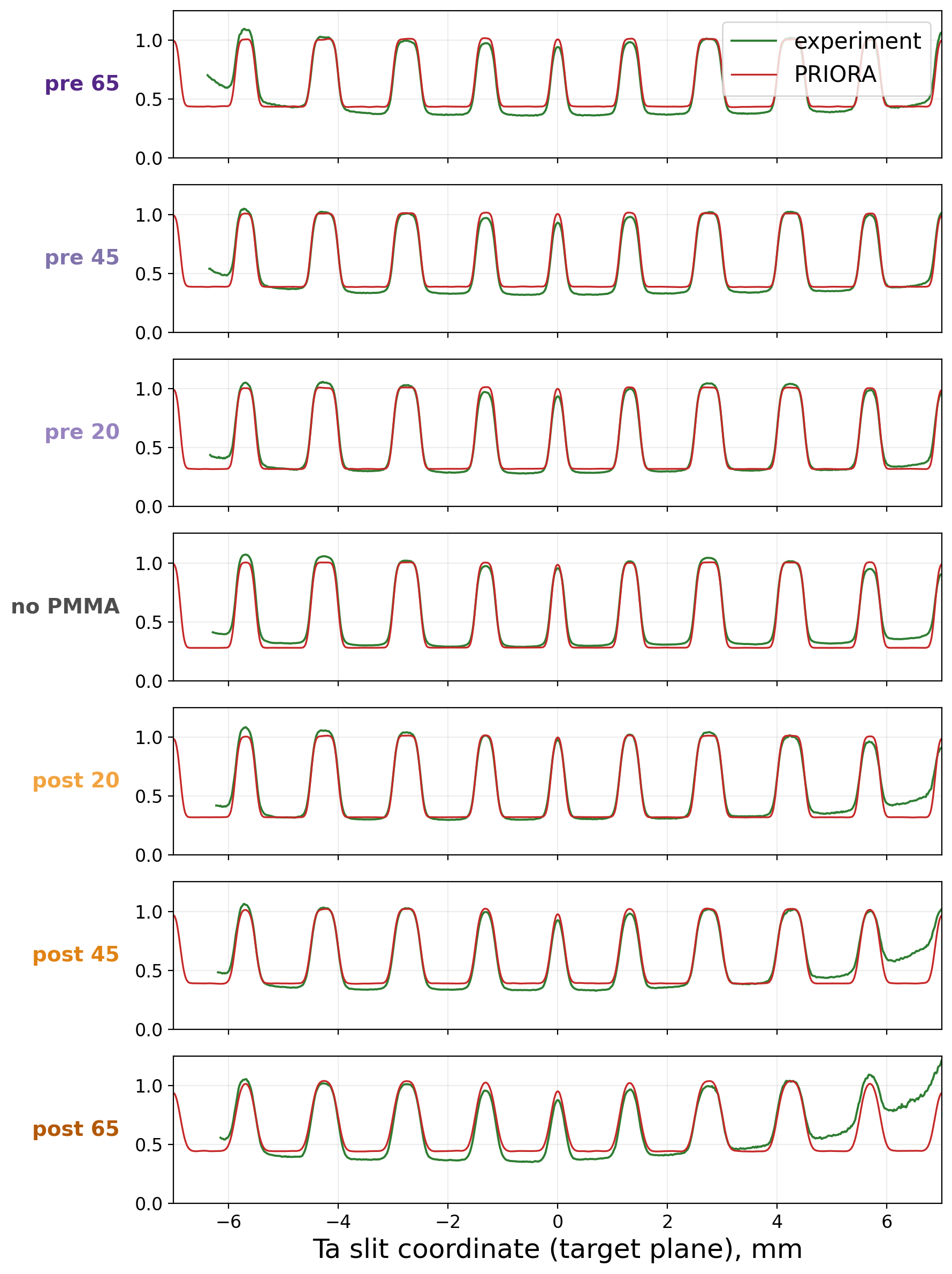}
  \caption{Band-wise transmission profiles of the 2\,GeV \tabench{}
  experiment (green) overlaid with the forward model of
  the reconstructed instrument state --- logged trims plus the
  $0.40$\,mm effective detector-response FWHM --- (red), in identical observables
  (Sec.~\ref{sec:conventions}); $N=10^{10}$ protons. The vignetted
  outermost slits ($\pm9.75$\,mm) are outside the plotted range.
  Band-label colours correspond to the profile-extraction regions in
  Fig.~\ref{fig:images}.}
  \label{fig:profiles}
\end{figure}

\begin{figure*}[tb]
  \centering
  \includegraphics[width=\textwidth]{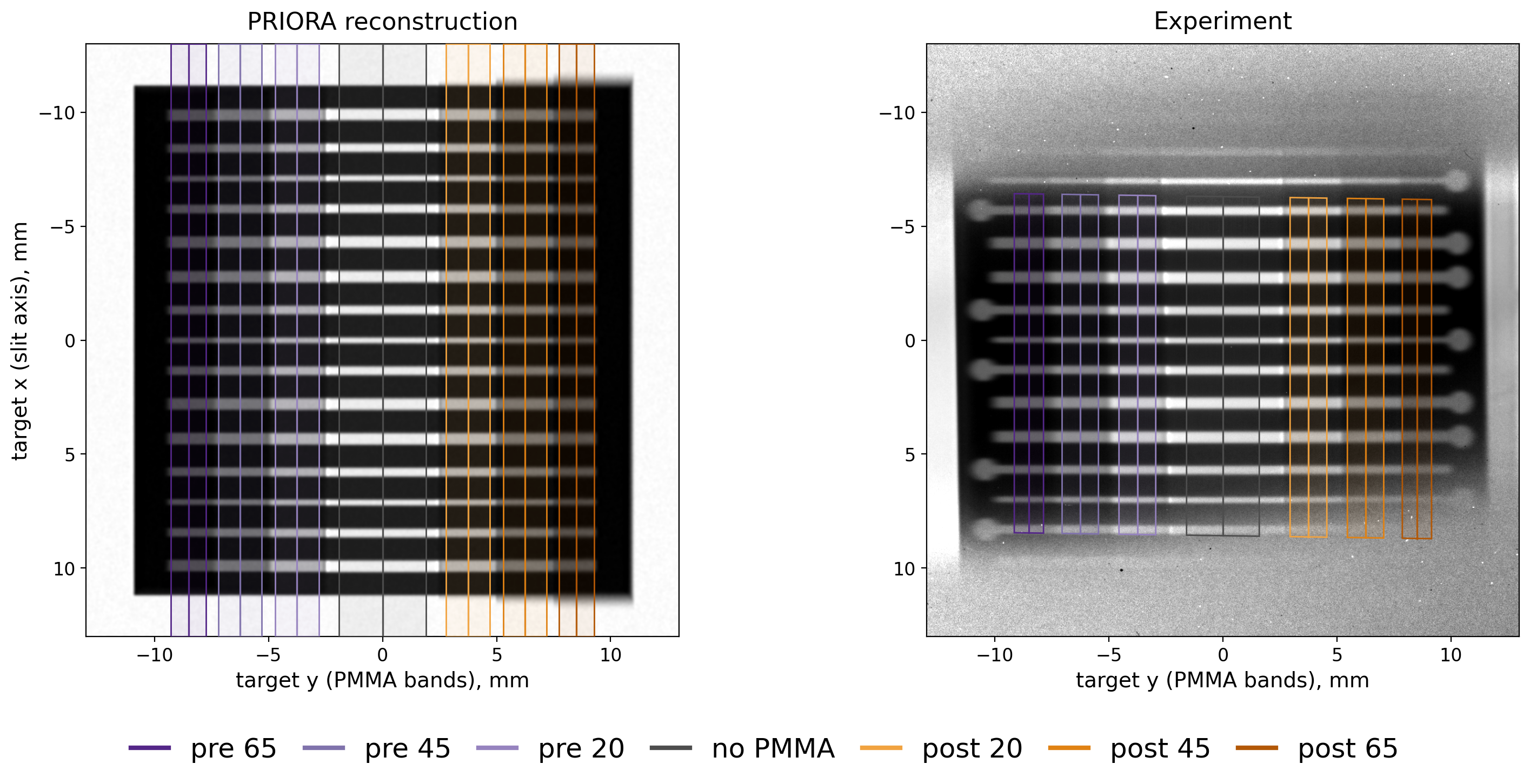}
  \caption{Synthetic radiograph of the reconstructed state (left) and
  the measured 2\,GeV image (right), in target-plane coordinates;
  $N=10^{10}$ protons. Both panels use the same target-plane grid and
  limits. Coloured rectangles mark the regions averaged to obtain the
  profiles in Fig.~\ref{fig:profiles}; the lines inside them mark the profile
  centres.}
  \label{fig:images}
\end{figure*}

Here the physics is fixed and the instrument state is fitted from one
radiograph. The design state gives ${\approx}26\,\mu$m ESF widths; the
data give 121--189\,$\mu$m. The PMMA-band dependence of the ESF widths
fixes the longitudinal focus sign. The no-PMMA and pre-PMMA widths are
nearly equal, whereas the post-PMMA widths grow with thickness. Therefore,
the slit-plane focus lies upstream of the Ta plate. PMMA step edges give
the orthogonal sign: pre-side boundaries are ${\approx}330\,\mu$m wide
versus ${\approx}100\,\mu$m post. That focus is downstream. The state is
astigmatic.

As a limiting case, the full no-PMMA width is attributed to optics.
This gives $dz_x\approx+30$\,mm. The resulting pure-defocus state predicts
6.5--9.1\%
overshoots or undershoots adjacent to the Ta edges, caused by chromatic
transport and the angular cut of edge-scattered protons. The corresponding
features in the measured profiles are only 0.2--4.8\%. Defocus alone
therefore cannot explain the measured width; an additional image-space
resolution term is required.

A joint fit of the profile shapes and ESF widths gives
$dz_x\approx+15$\,mm and an effective Gaussian detector-response FWHM of
$115\pm6\,\mu$m in the target plane ($0.40\pm0.02$\,mm at the detector).
This response is applied to the target and empty images before division and
therefore cannot create transmission contrast. It represents detector blur
and shot-to-shot beam and target motion in the multi-shot average. The 2 and
4\,GeV exposures therefore have independent response widths.

The experimental control-system logs, recovered after the PRIORA fit,
list quadrupole trims
$-0.71/+0.24/+0.48/-0.58\%$ (Q1--Q4). Their map gives
$dz_x=+15.6$\,mm, $dz_y=-67$\,mm and $\Delta M=+0.8\%$, consistent with
the measured magnification. Used without rescaling and combined with the
response, these trims reproduce all seven band widths with 7\% rms error,
the Ta-plate/open-slit transmission ratio, and the second-plane post-side
step width
(98 versus 97\,$\mu$m); Table~\ref{tab:bands},
Figs.~\ref{fig:profiles},~\ref{fig:images}. The same records give a
$54\times55$\,mm FWHM object illumination and matched correlation slopes.

\begin{table}[tb]
\caption{Band-wise slit ESF widths (FWHM, $\mu$m) of the 2\,GeV experiment
and of the forward model in the reconstructed state --- the logged
quadrupole trims plus the fitted $0.40$\,mm effective detector-response FWHM
($N=10^{10}$ protons per image) --- with the Ta-plate/open-slit
transmission ratio.}
\label{tab:bands}
\begin{ruledtabular}
\setlength{\tabcolsep}{4.5pt}
\begin{tabular}{lcccc}
Band & Exp. & Model & \shortstack{Exp.\\Ta/open} & \shortstack{Model\\Ta/open} \\
\hline
pre-PMMA 65  & 120.8 & 118.8 & 0.385 & 0.438 \\
pre-PMMA 45  & 123.7 & 120.8 & 0.344 & 0.387 \\
pre-PMMA 20  & 130.0 & 124.9 & 0.310 & 0.320 \\
no PMMA      & 122.0 & 130.5 & 0.315 & 0.280 \\
post-PMMA 20 & 145.7 & 134.0 & 0.322 & 0.321 \\
post-PMMA 45 & 166.5 & 166.4 & 0.362 & 0.390 \\
post-PMMA 65 & 188.8 & 217.6 & 0.412 & 0.447 \\
\end{tabular}
\end{ruledtabular}
\end{table}

The reconstruction is insensitive to the tested analysis and
beam-distribution variations. Fitted Ta-edge positions vary by only
${\approx}1\,\mu$m along each ROI. Increasing the initial momentum spread to
$5\times10^{-3}$ or varying the illumination correlations changes the
extracted widths within the Monte Carlo uncertainty. The remaining residuals
are the outer post-side 65\,mm band (+15\%) and the absolute pre-side
PMMA-step width in the orthogonal plane, likely caused by the mechanical
staircase joints.

\section{Cross-check at \texorpdfstring{4\,GeV}{4 GeV}}
\label{sec:4gev}

\begin{figure}[tb]
  \centering
  \includegraphics[width=\linewidth]{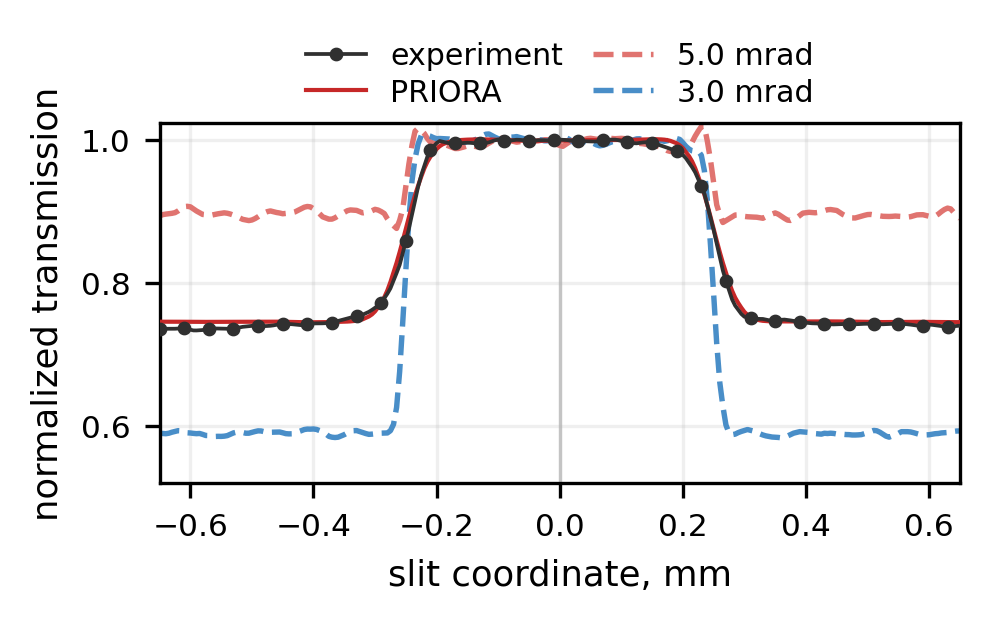}
  \caption{4\,GeV cross-check. Profiles of the four central 0.5\,mm Ta slits are
  translated to their fitted centres and averaged without rescaling;
  transmission is normalized to the open-slit level. The measured profile
  (points) is compared with PRIORA (red):
  $3.75$\,mrad effective acceptance, 69\,$\mu$m target-plane Gaussian exposure
  response and $N=10^{10}$ protons. Dashed curves show perfect-focus
  calculations for the nominal 3.0 and 5.0\,mrad apertures, which bracket the
  measured Ta/open transmission.}
  \label{fig:4gev}
\end{figure}

An independent experiment with the Ta-slit target was performed at
4\,GeV. The image shows none of the symmetric attenuation bands expected
from the mounted PMMA staircases (12, 7 and 3\%); the analysed
configuration is therefore the bare 2\,mm Ta plate. This removes the
chromatic-load constraints but leaves two direct measurements. First, the
ratio of transmission through Ta to that through an open slit is 0.743 in
the same central region used for the slit profiles.
Matching this ratio with PRIORA gives an effective angular acceptance of
3.75\,mrad, between the nominal 3.0 and 5.0\,mrad settings. This is the
acceptance transported through the complete Fourier-plane map, not a
direct measurement of the mechanical aperture. Second, the two-edge slit
fits give a target-plane spatial resolution of $69\pm8\,\mu$m FWHM. In the
zero-defocus model this corresponds to a 0.24\,mm effective Gaussian response
at the detector. Since the bare plate provides no longitudinal chromatic
lever, this response cannot be separated from a small focus error using this
image alone. The $N=10^{10}$ PRIORA calculation gives 0.746 for the Ta/open
transmission and 69\,$\mu$m for the spatial resolution
(Fig.~\ref{fig:4gev}).

The 2 and 4\,GeV measurements show the expected energy dependence. For
the same 2\,mm Ta plate, the scattering angle decreases from
$\theta_0\approx3.5$\,mrad at 2\,GeV to
$\theta_0\approx1.9$\,mrad at 4\,GeV. At 2\,GeV,
$\theta_0$ is comparable to the 3.0\,mrad acceptance, giving strong
contrast (Ta/open $\approx0.3$). At 4\,GeV, the effective
3.75\,mrad acceptance is larger than $\theta_0$, giving
weaker contrast (Ta/open $\approx0.74$) but a sharper measured edge.

\section{Conclusions}
\label{sec:conclusions}

The PRIORA forward model simulates proton transport through matter and a complete
ion-optical system by combining Monte Carlo matter physics with
arbitrary-order transfer maps; third-order energy-dependent maps are
used in this study. For the full PRIOR-II configuration, the calculation
rate on an 8-core laptop is approximately $10^6$ protons/s.

The \tabench{} target provides simultaneous tests of absolute scale,
angular acceptance, material contrast, spatial resolution and the
direction of longitudinal defocus in both transverse planes. A single 2\,GeV radiograph recovers
sub-percent quadrupole corrections; the same trims were later
cross-checked against the experimental logs. With this reconstructed
state, PRIORA describes all seven PMMA configurations with a
target-plane spatial resolution of $115\pm6\,\mu$m. The independent
4\,GeV experiment gives a central Ta/open transmission of 0.743 and a spatial
resolution of $69\pm8\,\mu$m; PRIORA gives 0.746 and
69\,$\mu$m, respectively.

The agreement in absolute transmission, spatial resolution and
chromatic-load dependence verifies the matter model and its coupling to
the ion optics. In particular, PRIORA recovers a non-nominal experimental
state instead of assuming the design tune. The comparison also resolves
the expected energy trade-off for the same Ta plate: increasing the
proton energy reduces scattering and material contrast while improving
spatial resolution. The PRIORA/\tabench{} target pair provides a baseline for
later designs and simulations and permits rapid tuning during experiments.

\begin{acknowledgments}
The authors thank the Plasma Physics Department of GSI
Helmholtzzentrum f\"ur Schwerionenforschung for support and for the
PRIOR-II experimental data used in this work.
\end{acknowledgments}

\end{document}